\documentclass[acmsmall,nonacm]{acmart}

\acmJournal{PACMPL}
\acmVolume{University at Buffalo, SUNY Technical Report}
\acmNumber{}
\acmArticle{}
\acmYear{2021}
\acmMonth{11}
\startPage{1}

\setcopyright{ccby}

\usepackage{amsmath}
\usepackage{amsthm}
\usepackage{bbm}
\usepackage{stmaryrd}
\usepackage{mathtools}
\usepackage{mathpartir}
\usepackage{pl-syntax}
\usepackage{doi}
\usepackage{ifthen}
\usepackage{xparse}
\usepackage{xspace}
\usepackage{xcolor}
\usepackage{tikz}
\usepackage{fontawesome}

\newcommand{\newcommenter}[3]{%
  \newcommand{#1}[1]{%
    \textcolor{#2}{\small\textsf{[#3: {##1}]}}%
  }%
}
\newcommenter{\akh}{purple}{AKH}
\newcommenter{\todo}{red}{TODO}

\theoremstyle{theorem}

\theoremstyle{definition}

\newcommand{\join}{\ensuremath{\mathbin{+\!\!+}}}

\newcommand{\believes}[2]{\ensuremath{\mathop{\square_{#1}} #2}}

\newcommand{\letl}[5]{\textsf{let}_{#1}\;#2.#3 \mathrel{\coloneqq} #4 \mathrel{\textsf{in}} #5}

\newcommand{\appfun}[2]{#1\;#2}

\newcommand{\send}[2]{#1 \mathbin{\leadsto} #2}

\newcommand{\upname}{\textsf{up}}
\newcommand{\up}[2]{\appfun{\upname_{#1}}{#2}}
\newcommand{\downname}{\textsf{down}}
\newcommand{\down}[2]{\appfun{\downname_{#1}}{#2}}

\newcommand{\unit}{\textsf{unit}\xspace}
\newcommand{\void}{\textsf{void}\xspace}

\newcommand{\lock}{\text{\faLock}}
\newcommand{\at}{\ensuremath{\mathrel{@}}}
\newcommand{\locks}[1]{\text{locks}(#1)}

\newcommand{\cansendname}{\text{CanSend}}
\newcommand{\cansend}[2]{\cansendname(#1, #2)}
\newcommand{\candownname}{\text{CanDown}}
\newcommand{\candown}[2]{\candownname(#1, #2)}
\newcommand{\canupname}{\text{CanUp}}
\newcommand{\canup}[2]{\canupname(#1, #2)}

\title[Corps]{{\small University at Buffalo Technical Report:}\\ Corps: A Core Calculus of Hierarchical Choreographic Programming}
\date{\today}

\author{Andrew K. Hirsch}
\orcid{0000-0003-2518-614X}
\affiliation{
  \institution{University at Buffalo}
  \city{Buffalo}
  \country{USA}
}
\email{akhirsch@buffalo.edu}

\begin{document}

% PACM Required stuff here
\begin{abstract}
  Functional choreographic programming suggests a new propositions-as-types paradigm might be possible.
In this new paradigm, communication is not modeled linearly; instead, ownership of a piece of data is modeled as a modality, and communication changes that modality.
However, we must find an appropriate modal logic for the other side of the propositions-as-types correspondence.
This paper argues for doxastic logics, or logics of belief.
In particular, authorization logics---doxastic logics with explicit communication---appear to represent \emph{hierarchical} choreographic programming.
This paper introduces hierarchical choreographic programming and presents Corps, a language for hierarchical choreographic programming with a propositions-as-types interpretation in authorization logic.
%%% Local Variables:
%%% mode: latex
%%% TeX-master: "conference"
%%% End:

\end{abstract}

\maketitle

\section{Introduction}
\label{sec:introduction}

Choreographic and multitier programming model concurrent systems as containing a set number of interacting processes.
Intuitively, these processes act as interacting agents, such as nodes in a distributed system.
However, propositions-as-types models of choreographies~\cite{CarboneMS14} do not respect this structure, instead exploiting the linear structure of communication implicit in choreographies.
The recent proliferation of \emph{functional} choreographic programming~\cite{HirschG22,GraversenHM23,CruzFilipeGLMP21}, however, suggests a different potential propositions-as-types interpretation: a non-linear logic with explicit agents.

\emph{Doxastic logic}, the logic of beliefs, seems like a natural fit.
Doxastic logics represent agents as modalities of belief.
If the agent~$A$ believes the formula~$\varphi$, then $\square_A \varphi$ holds.
Doxastic logics represent \emph{belief} rather than \emph{knowledge}, so $\square_A \varphi$ holding does not necessarily imply that $\varphi$ holds as this would imply that an agent is correct about all of their beliefs.
This is the key difference between doxastic and \emph{epistemic} logics.

In order for doxastic logics to serve as the other side of this propositions-as-types correspondence, we need to interpret formulae as types and proofs as data.
We propose that most connectives have their standard intuitionistic interpretations, while we interpret $\square_A \varphi$ as saying that the process~$A$ has a $\varphi$.
This fits with doxastic logic, rather than epistemic logic, as described above.

However, there one major component of doxastic logics does not fit naturally into the global concurrent programming paradigm: nested beliefs.
For instance, the formula $\mathop{\square_A} (\mathop{\square_B} \varphi)$ represents ``agent~$A$ believes that agent~$B$ believes~$\varphi$.''
Current functional choreographic languages do not allow nested claims of ownership over data.%
\footnote{PolyChor$\lambda$~\cite{GraversenHM23} uses a similar notion to represent delegation, though this does not seem to follow the logical laws of doxastic logic.
Meanwhile, in unpublished work \citeauthor{Kavvos24} has been exploring an interpretation of doxastic logic where agent~$A$ has a remote pointer to $B$'s data~\cite{Kavvos24}.}
For instance, in Pirouette, one of the first functional choreographic programming languages~\cite{HirschG22}, local data have types which are disjoint from choreographic data.
Indeed, it is not entirely clear what it would mean for agent~$A$ to have beliefs about agent~$B$.
However, note that because $\square_A$ represents a belief of $A$'s, there is no requirement that $A$'s beliefs about $B$ have any connection to $B$ at all.
Thus, $A$'s version of $B$ can be viewed as its own, independent agent.

With this view, agents are arranged in a forest topology, with each agent ``above'' its parent.
While this breaks with global-view tradition, where processes have a fully connected topology where anyone can talk to anyone else, this hierarchical topology for processes is far from unnatural.
For instance, in a distributed system where nodes can themselves be multithreaded, each node~$A$ can be represented by an agent, with each thread on~$A$ represented as an agent above $A$ in the tree.

It would not make sense for nodes to be unable to communicate with one another, so we would like to be able to communicate between processes on the same level.
However, no node~$A$ should be able to talk to another node~$B$'s thread without going through~$B$ itself.
We model this as allowing communication between sibling nodes.
This complicates the simple forest topology described earlier, since the children of every node form a fully connected topology.
In other words, each node can communicate with its parent, its children, \emph{and} its siblings.
We dub choreographic programming with this ``tree of interconnected processes'' topology \emph{hierarchical choreographic programming}.

This leads to \emph{authorization logic}, a version of doxastic logic with explicit communication.
However, authorization logic allows for more-subtle topologies by restricting which siblings may communicate.
In particular, authorization logic contains a new judgment---$A \mathrel{\textsf{speaksfor}} B$---representing the ability of agents~$A$ and~$B$, who are at the same level in the tree, to communicate.
Then, if $\square_A \varphi$ and $A \mathrel{\textsf{speaksfor}} B$, then $A$ can send its proof of $\varphi$ to $B$, allowing us to derive $\square_B \varphi$.
We can thus think of \textsf{speaksfor} as a one-way channel---since $A$ can send to $B$, but not vice versa---and \textsf{speaksfor}-elimination as the choreographic ``send-and-receive'' primitive.

This paper describes ongoing research into formalizing this connection between doxastic logics, especially authorization logics, and global concurrent programming.
First, we describe Corps, a core calculus for hierarchical choreographic programming, allowing us to form the tree topology described above with explicit intra-level communication.
We then provide a number of conjectures about Corps and its relationships to doxastic and authorization logics.
By providing an explicit propositions-as-types interpretation of authorization logic in choreographies, we hope to be able to advance both technologies considerably.

%%% Local Variables:
%%% mode: latex
%%% TeX-master: "conference"
%%% End:

\section{Background: Doxastic and Authorization Logics}
\label{sec:backgr-doxast-auth}

As described in Section~\ref{sec:introduction}, doxastic logics are modal logics of belief.
In this section, we present the main rules of a simple authorization logic in Fitch style~\cite{Clouston18}.
One rule requires reasoning about trust (modeled as a relation called \textsf{speaksfor}) and communication directly.
This rule separates authorization logics from other doxastic logics.

\begin{figure}
  \centering
  \begin{syntax}
    \abstractCategory[Agents]{A, B}
    \category[Generalized Agents]{g} \alternative{\diamond} \alternative{g \cdot A}
    \category[Formulae]{\varphi, \psi}
      \alternative{\top}
      \alternative{\bot}
      \alternative{\varphi \land \psi}
      \alternative{\varphi \lor \psi}
      \alternative{\varphi \to \psi}
      \alternative{\believes{A}{\varphi}}
    \category[Contexts]{\Gamma, \Delta}
      \alternative{\cdot}
      \alternative{\Gamma, \varphi \at g}
      \alternative{\Gamma, \lock_g}
  \end{syntax}
  
  \caption{Syntax of Doxastic Logic}
  \label{fig:dox-syn}
\end{figure}

The syntax of doxastic logic formulae can be found in Figure~\ref{fig:dox-syn}.
The syntax of the logic is parameterized on a set of \emph{agents} about which we wish to reason; we denote elements of this set with capital Latin characters like~$A$.
The formulae (denoted with lower-case Greek letters like~$\varphi$) mostly consist of those of intuitionistic propositional logic, with their usual meanings.
However, we also have the \emph{modality} $\believes{A}{}$, representing $A$'s beliefs.
Thus, we should read $\believes{A}{\varphi}$ as ``$A$ believes $\varphi$.''

The contexts (denoted by capital Greek letters like~$\Gamma$), however, are more unusual.
Most logics use lists of formulae for their contexts, and then use structural rules (or their equivalent) to force those lists to act as sets.
Here, we use sets of formulae tagged with \emph{generalized agents}, separated by similarly-tagged locks.
Intuitively, a generalized agent $g$ represents beliefs-of-beliefs: the generalized agent $\diamond$ is ground truth, $\diamond \cdot A$ is the agent $A$, $\diamond \cdot A \cdot B$ is $A$'s beliefs about $B$, and so on.
Thus, a generalized agent represents a path up the tree topology described in Section~\ref{sec:introduction}.
Then $\varphi \at g$ represents that $g$ believes $\varphi$---$\!\!{}\at \diamond \cdot A$ internalizes $\believes{A}{}$ in the same way that commas internalize conjunctions.
The context $\Gamma, \lock_g$ represents $g$'s view of the context $\Gamma$---$g$ can access its own beliefs, but not the beliefs of others.

Generalized agents are essentially (snoc) lists of agents, and thus form a monoid in a standard way.
We write the monoid action as $\join$.
In particular, $$g \join g' = \left\{\begin{array}{ll} g & \text{if $g' = \diamond$}\\ (g \join g'') \cdot A & \text{if $g' = g'' \cdot A$}\end{array}\right.$$
We can then compute the locks in a context.
We consider a judgment $\Gamma \vdash \varphi$ to be a proof of $\varphi$ ``from $\locks{\Gamma}$'s point of view.''
$$\locks{\Delta} = \left\{\begin{array}{ll} \diamond & \text{if $\Delta = \cdot$}\\ \locks{\Delta'} & \text{if $\Delta = \Delta', \varphi \at g$}\\ \locks{\Delta'} \join g & \text{if $\Delta = \Delta', \lock_g$}\end{array}\right.$$
Contexts also use the monoid action freely to reason about locks; in other words, contexts obey the equations $\Gamma, \lock_\diamond = \Gamma$ and $\Gamma, \lock_{g_1}, \lock_{g_2} = \Gamma, \lock_{g_1 \join g_2}$.

\begin{figure}
  \centering

  \begin{mathpar}
    \infer*[left=Axiom]{\locks{\Delta} = g }{\Gamma, \varphi \at g, \Delta \vdash \varphi} \\
    \infer*[left=BelievesI]{\Gamma, \lock_A \vdash \varphi}{\Gamma \vdash \believes{A}{\varphi}}\and
    \infer*[right=BelievesE]{\Gamma, \lock_{g_1} \vdash \believes{g_2}{\varphi} \\ \Gamma, \varphi \at g_1 \mathbin{+\!\!+} g_2 \vdash \psi}{\Gamma \vdash \psi} \and
    \infer*[left=SelfDown]{\Gamma \vdash \believes{A}{(\believes{A}{\varphi})}}{\Gamma \vdash \believes{A}{\varphi}} \and
    \infer*[right=SelfUp]{\Gamma \vdash \believes{A}{\varphi}}{\Gamma \vdash \believes{A}{(\believes{A}{\varphi})}}\\
    \infer*[right=SpeaksforE]{A \mathrel{\textsf{speaksfor}} B\\ \Gamma \vdash \believes{A}{\varphi}}{\Gamma \vdash \believes{B}{\varphi}}
  \end{mathpar}
  
  \caption{Selected Rules for Doxastic Logic and Authorization Logic}
  \label{fig:dox-rules}
\end{figure}

Figure~\ref{fig:dox-rules} contains selected rules of authorization logic.
(The rule \textsc{SpeaksforE} is a rule of authorization logic not found in standard doxastic logic.)
The rules presented here define the behavior of the $\believes{A}{}$ modality.
Only the standard rules of intuitionistic propositional logic remain.

The first rule is \textsc{Axiom}, which allows the use of an assumption~$\varphi \at g$.
In order to use such an assumption, we must know that the locks to the left of the assumption form $g$.
In other words, we must know that the proof is ``from $g$'s point of view,'' and thus it is appropriate to assume $\varphi$ from the fact that $g$ believes it.

The next two rules are the introduction and elimination rules for the $\believes{A}{}$ modality.
The first says that we can prove $\believes{A}{\varphi}$ if we can prove $\varphi$ with an $A$-labeled lock.
Since we can view locking a context as taking on $A$'s point of view, this says that in order to prove that $A$ believes $\varphi$, we can prove $\varphi$ from $A$'s point of view.
The elimination rule is more complicated.
Ignoring the role of~$g_1$ for the moment, \textsc{BelievesE} says that if you can prove $\believes{g_2}{\varphi}$, you may use the assumption $\varphi \at g_2$ to prove $\psi$.
Here, we're using a new piece of shorthand: $\believes{g}{\varphi}$ stands for a stack of modalities on top of $\varphi$, one for each part of $g$.
Thus, this says that if you can prove that $g_2$ believes $\varphi$, and you can prove $\psi$ assuming that $g_2$ believes $\varphi$, then you can prove $\psi$ immediately.
Adding in $g_1$ allows for more-flexible use of assumptions in the proof of $\believes{g_2}{\varphi}$, but necessitates adding that flexibility as requirement on the proof of $\psi$.

Finally, the last two rules say that $A$ has perfect knowledge about their own beliefs.
First, \textsc{SelfDown} says that if $A$ believes that $A$ itself believes something, it is true.
Second, \textsc{SelfUp} says that if $A$ believes something, then they believe that they believe it.
In the context of beliefs, these both make perfect sense; as we will see in Section~\ref{sec:lang-hier-chor}, we will need to generalize this in order to match many communication topologies which we care about in practice.

\paragraph{From Doxastic to Authorization Logic}
Authorization logics are doxastic logics with a notion of trust and communication among trusted lines.
For our purposes, authorization logic is parameterized by an additional preorder between agents, called \textsf{speaksfor}.
Then, we add one rule to our doxastic logic, called \textsc{SpeaksforE} in Figure~\ref{fig:dox-rules}.
This rule says that if $A$ speaks for $B$, and $A$ believes $\varphi$ (often written $A \mathrel{\textsf{says}} \varphi$ in authorization logics), then $B$ also believes $\varphi$.
Intuitively, if $A$ speaks for $B$, then $A$ is willing to send its evidence for $\varphi$ to $B$, and $B$ is willing to incorporate $A$'s evidence into its own beliefs.
Thus, we can think of \textsc{SpeaksforE} as representing a message send in a concurrent system, very similar to the choreographic send-and-receive operation.

%%% Local Variables:
%%% mode: LaTeX
%%% TeX-master: "conference"
%%% End:

\section{A Language for Hierarchical Choreographic Programming}
\label{sec:lang-hier-chor}

\begin{figure}
  \centering
  \begin{syntax}
    \category[Types]{\tau}
    \alternative{\unit}
    \alternative{\void}
    \alternative{\believes{A}{\tau}}
    \alternative{\tau_1 \times \tau_2}
    \alternative{\tau_1 + \tau_2}
    \alternative{\tau_1 \to \tau_2}
    \category[Expressions]{e}
    \alternative{x}
    \alternative{A.e}
    \alternative{\letl{g_1}{g_2}{x}{e_1}{e_2}}\\
    \alternative{\send{e}{A}}
    \alternative{\up{g}{e}}
    \alternative{\down{g}{e}}
  \end{syntax}
  \begin{mathpar}
    \infer*[left=Axiom]{
      \locks{\Delta} = g
    } {
      \Gamma, x : \tau @ g, \Delta \vdash x : \tau
    }\\
    \infer*[left=BelievesI]{
      \Gamma, \lock_A \vdash e : \tau
    }{
      \Gamma \vdash A.e : \believes{A}{\tau}
    }\and
    \infer*[right=BelievesE]{
      \Gamma, \lock_{g_1} \vdash e_1 : \believes{g_2}{\tau}\\
      \Gamma, x : \tau \at g_1 \join g_2 \vdash e_2 : \sigma
    }{
      \Gamma \vdash \letl{g_1}{g_2}{x}{e_1}{e_2} : \sigma
    }\\
    \infer*[right=Down]{
      \Gamma \vdash e : \believes{g}{\tau}\\
      \candown{\locks{\Gamma}}{g}
    }{
      \Gamma \vdash \down{g}{e} : \tau
    }\and
    \infer*[left=Up]{
      \Gamma \vdash e : \tau\\
      \canup{\locks{\Gamma}}{g}
    }{
      \Gamma \vdash \up{g}{e} : \believes{g}{\tau}
    }\\
    \infer*[left=Send]{
      \Gamma \vdash e : \believes{g_1}{\tau}\\
      \cansend{g_1}{g_2}
    }{
      \Gamma \vdash \send{e}{g_2} : \believes{g_2}{\tau}
    }\and
  \end{mathpar}
  \caption{Corp Selected Syntax and Typing Rules}
  \label{fig:choreo-types}
\end{figure}

So far, we have focused on the purely logical presentation of doxastic and authorization logics.
We now present Corps (short for ``Corps de Ballet,'' a rank in ballet troupes), a language for \emph{hierarchical choreographic programming}.
The type syntax, along with selected pieces of the program syntax and the corresponding typing rules, can be found in Figure~\ref{fig:choreo-types}.
Every type connective has its standard interpretation in logic.
For instance, the type connective~$\times$ corresponds to the logical connective~$\land$.
The only exception is $\believes{A}{\tau}$, which represents a $\tau$ stored on the agent~$A$ located one level up in the hierarchy.

We can view a typing judgment $\Gamma \vdash e : \tau$ as saying that $e$ computes a $\tau$ on the process $\locks{\Gamma}$.
From this perspective, the \textsc{Axiom} rule says that we can use $x$ as a $\tau$ if the current process has access to a $\tau$ named $x$ in the context.
Furthermore, the \textsc{BelievesI}~rule says that to compute a $\believes{A}{\tau}$ on the process $\locks{\Gamma}$, we need to compute a $\tau$ on $\locks{\Gamma} \cdot A$.
The \textsc{BelievesE}~rule says that if we can compute a $\believes{g_2}{\tau}$ on the process $g_1$, then we can use the result of that computation as a $\tau$ on the process $g_1 \join g_2$.

The remaining rules correspond not only to similar rules in logic, but to explicit communication.
The rule~\textsc{Down} corresponds to the rule~\textsc{SelfDown} in doxastic logic.
However, as we can see here, it is more general than the original \textsc{SelfDown}~rule.
Just as authorization logic is parameterized by the \textsf{speaksfor} relation between agents to determine when they can communicate, Corps is parameterized by three similar relations between generalized agents.
Here, the relation~$\candownname$ determines when communication is allowed \emph{down the tree}.
To recover the \textsc{SelfDown}~rule, we can set $\candown{g_1}{g_2} \overset{\Delta}{\iff} g_1 = g \cdot A \land g_2 = g \cdot A \cdot A$; that is, $g_2$ is allowed to communicate down to $g_1$ only if $g_1$ is some agent~$A$ (possibly up the tree by some path), and $g_2$ is the agent named~$A$ directly above $g_1$.
This allows communication down the tree only from $A$'s beliefs about itself.
Similarly, the rule~\textsc{Up} corresponds to the original \textsc{SelfUp}~rule, but generalized with the $\canupname$~relation controlling when communication is allowed up the tree.
Finally, the rule~\textsc{Send} corresponds directly to the original \textsc{SpeaksforE}~rule, with \cansendname{} corresponding to \textsf{speaksfor}.

By setting different meanings for \candownname{}, \canupname{}, and \cansendname{}, we can get very different topologies all within the hierarchical style.
For instance, to get the traditional fully-connected topology between sibling nodes for choreographies, we need only set $\cansend{g_1}{g_2} \overset{\Delta}{\iff} \textsf{True}$.
Many of the traditional settings for these relations only involve a postfix of the generalized agent, similar to the traditional setting for \candownname{} discussed earlier.
This creates a uniform set of policies: every agent allows its children of the same name to communicate both with itself and with each other.
However, the generality of these relations allows for different (generalized) agents to have different policies.
We expect that many topologies that arise in practice will not match the requirement that a node named~$A$ only communicate with a thread also named~$A$, leading to this generalization.

%%% Local Variables:
%%% mode: LaTeX
%%% TeX-master: "conference"
%%% End:

\section{Conjectures}
\label{sec:conjectures}

In current work we are exploring endpoint projection for Corps.
While in previous works processes were in one-to-one correspondence with agents, in Corps processes correspond to \emph{generalized} agents.
Thus, each node at every level of the tree is a process.
Endpoint projection is then a type-directed translation $\llbracket \Gamma \vdash e : \tau \rrbracket_{g}$, which is non-trivial only when $g$ is at or above $\locks{\Gamma}$.

We are also currently developing a normalizing semantics for Corps, in the usual sense of $\lambda$-calculus.
However, the connection to logic suggests that there are two notions of normalization that will be interesting to study: one which does not reduce any communications, and one which reduces communications of positive forms.
The first corresponds to cut elimination in sequent-calculus presentations of authorization logic, while the second corresponds to a stronger normal form of proofs where only variables may be communicated.
Both \citet{HirschACAT20} and \citet{GratzerNKB20} have found that this stronger normal form is useful when studying modal logics with communication; this suggests a deep reason why.
We plan to use a labeled-transition semantics of Corps which reduces communication to show that authorization and doxastic proofs enjoy this stronger normalization property.
The correctness of endpoint projection and normalization also suggest that Corps is deadlock free.

In authorization logics, an important property is \emph{noninterference}: if $A \mathrel{\textsf{speaksfor}} B$ is \emph{not} true, then $A$'s beliefs cannot affect $B$'s beliefs~\cite{HirschACAT20,HirschC13,GargP06,JiaVMZZSZ08}.
We believe a similar property is true for Corps, taking into account the \canupname{} and \candownname{}~relations in addition to \cansendname{}.
However, we believe that this will be significantly easier to prove for Corps, taking advantage of proof techniques for noninterference in information-flow control systems~\cite{MenzHLG23,SilverHCHZ23,HirschC21}.
We hope to leverage these proof techniques to allow for noninterference proofs for more complicated authorization logics in the future, as previous attempts have been stymied by the complex proof-theoretic constructions involved.

%%% Local Variables:
%%% mode: LaTeX
%%% TeX-master: "conference"
%%% End:

\section{Conclusion}
\label{sec:conclusion}

In this work, we presented Corps, a language for \emph{hierarchical choreographic programming} with a propositions-as-types connection to authorization logic and doxastic logics more broadly.
While Corps is still very much an early work in progress, we believe that there are significant early results suggesting the fundamental nature of this work.
Moreover, should our conjectures hold, we can use them to translate between results about the type theory of functional choreographic programs and the proof theory of authorization logic.
Since the complexity of authorization-logic proof theory has prevented much development in the field, we are hopeful that this will spurn much development in both fields.

We hope that this work will inspire changes in how people design functional choreographic programming languages.
When restricted to only ``first-tier'' agents, Corps programs are close to  other functional choreographic programming languages, like Pirouette~\cite{HirschG22} and Chor$\lambda$~\cite{GraversenHM23,CruzFilipeGLMP21}.
However, subtle differences coming from the logical side have begun to appear; by exploiting those, we believe that we can design more grounded, foundational choreographic languages.

Finally, we note that this work suggests a deep connection between linear and doxastic logics: since deadlock-free communication seems to spring from both linear logic (via session types) and doxastic logics (via hierarchical choreographic programming).
The nature of this connection is unclear; each proof in doxastic logic seems to give rise to a series of linear-logic proofs, one for each projected program.
However, given that no-one has as of yet been able to show that projections of choreographic programs follow a linear-logic session type, this is unclear.
Nevertheless, tugging on this thread may lead to ideas that will change our understanding of both logic and concurrency.

%%% Local Variables:
%%% mode: latex
%%% TeX-master: "conference"
%%% End:

\bibliography{main}  
\end{document}